\documentclass[twocolumn]{jpsj2} %% two-column layout
\title{Influence of Homeotropic Anchoring Walls upon Nematic and
Smectic Phases} 
\author{
Masashi  \textsc{Torikai}\thanks{E-mail address: torikai@phen.mie-u.ac.jp}
and
Mamoru \textsc{Yamashita}\thanks{E-mail address: mayam@phen.mie-u.ac.jp}}

\inst{Department of Physics Engineering, Faculty of Engineering, Mie
University, 1515 Kamihama, Tsu 514-8507}

\abst{
McMillan liquid crystal model sandwiched between strong homeotropic
anchoring walls is studied.
Phase transitions between isotropic, nematic, and smectic A phases are
investigated for wide ranges of an interaction parameter and of the
system thickness.
It is confirmed that the anchoring walls induce an increase in
transition temperatures, dissappearance of phase transitions, and
an appearance of non-spontaneous nematic phase.
The similarity between influence of anchoring walls and that of external
fields is discussed. 
}

\kword{liquid crystal, McMillan model, homeotropic anchoring, nematic
phase, smectic A phase, non-spontaneous nematic phase}

\begin{document}
\maketitle

\section{Introduction}
Liquid crystals are very sensitive to conditions of boundary
surfaces.
The anchoring conditions at the boundary walls strongly influence the
liquid crystalline orderings if the systems are thin enough.
The isotropic(I)-nematic(N) phase transition is easily affected by the 
anchoring walls since the discontinuity at the I-N phase transition is
small. 
It is known that a thin nematogen system sandwiched between
homeotropic anchoring walls exhibits the I-N transition at higher
temperature than the I-N transition temperature of the system without
boundary anchoring~\cite{Sheng1976,Yokoyama1988,Poniewierski1987}.
As the system becomes thin, the discontinuity at the I-N transition
decreases;
eventually the I-N transition vanishes if the system is thinner than a
critical thickness~\cite{Sheng1976}. 
It is also shown theoretically that, even if the system exhibits
I-smectic A(A) phase transition in bulk systems(i.e., in the absence of
the walls), the homeotropic anchoring walls can induce a surface N
phase~\cite{Lelidis2002}; 
such a N phase is called a non-spontaneous N phase.
Corresponding to this result, strong N order induced near the anchoring
walls has found in decylcyanobiphenyl (10CB), which exhibits direct I-A
transition in bulk systems~\cite{Moses2001}.
The fact that such a wall-induced N order has not found in
undecylcyanobiphenyl (11CB) and dodecylcyanobiphenyl (12CB) indicates
that 10CB has stronger tendency to exhibit the N phase than others.
Such a tendency is to be expected since the 10CB is the shortest
molecule among the homologous series of cyanobiphenyl $n$CB exhibiting
direct I-A transition in bulk systems.

Although studies on a homologous series would provide a unified
understanding of the phase behavior, there are no theoretical studies on
the influence of anchoring walls upon members of homologous series.
In the present paper, we investigate such influence in systematic way;
in order to do so, we use a so-called McMillan model sandwiched between
strong homeotropic anchoring walls.
The McMillan model can mimic several materials of a homologous series
by choosing a parameter $\alpha$, contained in the model;
depending on the $\alpha$, the McMillan model exhibits three phases (I,
N and A) or two phases (I and A).
We report the influence of the anchoring walls, i.e., the changes in
phase transition temperatures, the critical thicknesses and their
dependence on $\alpha$, and the non-spontaneous N phase induced by
anchoring walls.
In the last section, we will discuss the similarity between the
influence of anchoring walls and of the external fields on the McMillan
model. 

\section{Discrete McMillan Model}
Self-consistent equations for the discrete McMillan model in
inhomogeneous systems have been derived in ref.~\citen{Torikai2004}, as
a generalization of the original McMillan model~\cite{McMillan}.
Thus in this section we will shortly give definitions and formulae.

%1
The N and A phase order parameters are functions of the distance
from one of walls in this system.
We divide the system into layers parallel to the walls and assume the
order parameters within each layer are constant.
Let the number of layers be $N+2$, and zeroth and $(N+1)$-th layers be
in contact with walls.
In the discrete McMillan model, the number of layers $N$ can be
interpreted as the system thickness.
The N order parameter $s_{n}$ and A order parameter $\sigma_{n}$ of
$n$-th ($0 \leq n \leq N+1$) layer are, by assuming that the $z$-axis is
parallel to the director, defined as 
\begin{subequations}
 \begin{align}
  s_{n} =
  & \left\langle P_{2}(\cos \theta)\right\rangle_{n}
  \label{eq:defOfNorder},\\
  \sigma_{n} =
  & \left\langle
  \cos \left(\frac{2 \pi}{d} z \right) P_{2}(\cos \theta) 
  \right\rangle_{n}
  \label{eq:defOfAorder},
 \end{align}\label{eq:defOfOrder}
\end{subequations}
where $P_{2}(x)=(3x^{2}-1)/2$;
$\theta$ and $z$ are the polar angle and $z$-coordinate of a molecule,
respectively. 
The brackets $\langle \cdot \rangle_{n}$ denote the thermal average
over molecules in $n$-th layer.

%2
We take into account only the molecular fields produced by neighboring
molecules. 
Then the one-particle potential for a molecule in the $n$-th layer is
%\begin{equation}
% v_{n} =
% -[V s_{n} + \tilde{h}_{s}^{(n)}] P_{2}(\cos \theta)
% -[\alpha V \sigma_{n} + \tilde{h}_{\sigma}^{(n)}]
% \cos\left(\frac{2 \pi}{d} z \right) P_{2}(\cos \theta)
% \label{eq:oneParticlePotential},
%\end{equation}
\begin{multline}
 v_{n} =
 -[V s_{n} + \tilde{h}_{s}^{(n)}] P_{2}(\cos \theta)\\
 -[\alpha V \sigma_{n} + \tilde{h}_{\sigma}^{(n)}]
 \cos\left(\frac{2 \pi}{d} z \right) P_{2}(\cos \theta)
 \label{eq:oneParticlePotential},
\end{multline}
where $Vs_{n}$ and $\alpha V \sigma_{n}$ are molecular fields of the
$n$-th layer;
$\tilde{h}_{s}^{(n)} = V'(s_{n-1} + s_{n+1} - 2s_{n})$ and
$\tilde{h}_{\sigma}^{(n)} = \alpha V'(\sigma_{n-1} + \sigma_{n+1} - 2\sigma_{n})$
are molecular fields of neighboring layers, that vanish in 
homogeneous systems.
The parameter $\alpha$ is an interaction strength for A phase;
$\alpha$ can be assumed to be an increasing function of the molecular
length~\cite{McMillan}. 
Using the discrete McMillan model potential
\eqref{eq:oneParticlePotential}, we obtain the self-consistent
equations: 
\begin{subequations}
 \begin{align}
  s_{n}&= 
  I(\beta (V s_{n} + \tilde{h}_{s}^{(n)}),
  \beta (\alpha V \sigma_{n} + \tilde{h}_{\sigma}^{(n)}))
  \label{eq:scEqForN},\\
  \sigma_{n}&=
  J(\beta (V s_{n} + \tilde{h}_{s}^{(n)}),
  \beta (\alpha V \sigma_{n} + \tilde{h}_{\sigma}^{(n)}))
  \label{eq:scEqForA},
 \end{align}\label{eq:scEq}
\end{subequations}
where functions $I$ and $J$ are defined as
\begin{subequations}
 \begin{align}
  I(\eta, \zeta)=\frac{\partial}{\partial \eta}
  \ln Z(\eta, \zeta),
  \\
  J(\eta, \zeta)=\frac{\partial}{\partial \zeta}
  \ln Z(\eta, \zeta),
 \end{align}\label{eq:IandJ}
\end{subequations}
where
%\begin{equation}
% Z(\eta, \zeta) =
%  \int_{0}^{\pi} d\theta \sin \theta \int_{0}^{d} dz
%  \exp\left[
%       \left\{
%        \eta + \zeta \cos\left(\frac{2 \pi}{d} z \right)
%       \right\}
%       P_{2}(\cos \theta)
%      \right]. \label{eq:partitionFunction}
%\end{equation}
\begin{multline}
 Z(\eta, \zeta) =
  \int_{0}^{\pi} d\theta \sin \theta \int_{0}^{d} dz \\
 \times \exp\left[
       \left\{
        \eta + \zeta \cos\left(\frac{2 \pi}{d} z \right)
       \right\}
       P_{2}(\cos \theta)
      \right]. \label{eq:partitionFunction}
\end{multline}

%2-4
Among the sets of solutions of eqs.~\eqref{eq:scEq}, the thermodynamic
stable set gives the minimum of a function
\begin{multline}
 \beta
 \tilde{F}(\beta, \{s_{n}, \sigma_{n}\})
 =
 \sum_{n=1}^{N}
 \left\{
 \frac{\beta V}{2} s_{n}{}^{2}
 +\alpha \frac{\beta V}{2} \sigma_{n}{}^{2} \right. \\
 \left.
 -\ln \frac{Z(\beta(V s_{n} + \tilde{h}_{s}^{(n)}),
 \beta (\alpha V \sigma_{n} + \tilde{h}_{\sigma}^{(n)})}
 {Z(0, 0)}
 \right\}  \\
 + \sum_{n=1}^{N-1}
 \left\{
 V' s_{n} s_{n+1} +
 \alpha V'  \sigma_{n} \sigma_{n+1}
 \right\}
 ,\label{eq:FforThin}
\end{multline}
where the set $\{s_{n}, \sigma_{n}\}$ is one of solutions to
eqs.~\eqref{eq:scEq}.
The minimum of $\tilde{F}$ is the thermodynamic free energy.

\section{Results}
For later discussions, it is suitable here to refer to the phase diagram
of the McMillan model without boundaries~\cite{McMillan}.
This model exhibits some different behaviors depending on the value of
$\alpha$.
For $\alpha \geq 0.98$, A phase directly melts into I phase;
this transition is first-order.
The N phase appears as an intermediate phase between I and A phases for
$\alpha \leq 0.98$.
The I-N transition temperature for $\alpha \leq 0.98$ is
$T=0.2202$ (in units of $V$), which does not depend on $\alpha$;
the triple point of I, N and A phase is thus $\alpha_\text{triple}=0.98$
and $T_\text{triple}=0.2202$. 
The I-N phase transition is first-order for any $\alpha$, while the N-A
phase transition is first-order for $0.70 \leq \alpha \leq 0.98$ and
second-order for $\alpha \leq 0.70$;
i.e., there is a tricritical point on the N-A coexisting line (at
$\alpha_\text{tc}=0.70$ and $T_\text{tc}=0.1910$).

In the presence of walls or fields, the N order is induced even if the
temperature is high;
such a high temperature phase with finite N order is called a
para-nematic phase. 
However, for simplicity, we do not use the term para-nematic but use the
terms of systems without walls and fields;
we call the high temperature phase I phase, the low temperature phase A
phase, and the intermediate phase N phase, if it exists.

In this work, we assume the homeotropic anchoring condition,
i.e., the director is perpendicular to the walls.
In addition, we assume that the anchoring is so strong that the order
parameters at the walls are extremely large.
Thus we will solve the self-consistent equations under the boundary
conditions $s_{0}=s_{N+1}=1$.
Furthermore, we also assume that the strong A ordering is induced at the
walls, i.e., $\sigma_{0}=\sigma_{N+1}=1$. 
We set the potential parameters $V=1$ and $V'=V/6$.

\begin{figure}
 \begin{center}
  \includegraphics[width=8.5cm,clip]{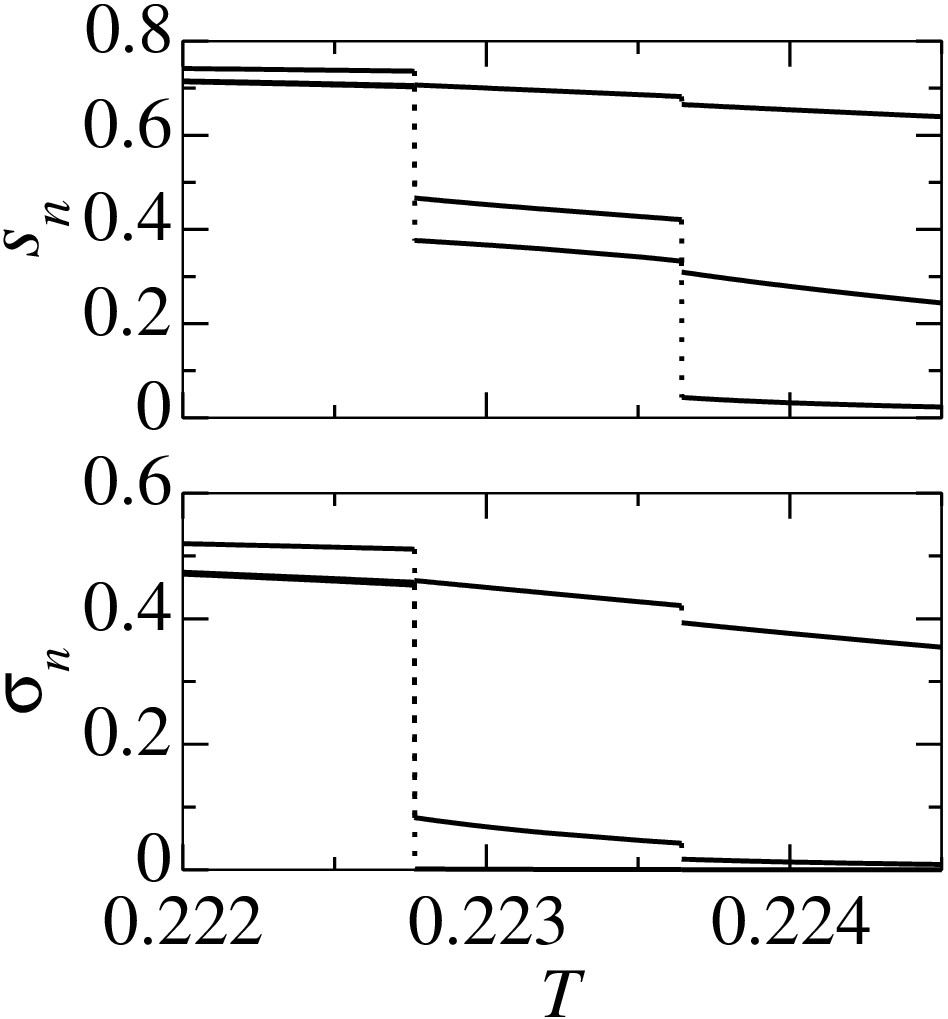}
 \end{center}
 \caption{
 Temperature dependence of order parameters of the $n$-th layer in a
 system of $N=18$ and $\alpha=1.0$.
 From top to bottom, curves are for layers $n=2$, $5$, and $9$.
 }
 \label{fig:snAndSigman}
\end{figure}

A typical behavior of order parameters of layers is shown in
fig.~\ref{fig:snAndSigman} for a system of $\alpha=1.0$ and thickness
$N=18$.
In our calculations, as shown in fig.~\ref{fig:snAndSigman}, all the
layers undergo each phase transition at the same temperature,
i.e., the transition temperatures do not depend on the distance from the
walls.
Thus in order to find each transition temperature we can use averaged
order parameters, 
$\langle s \rangle =\sum_{n} s_{n}/N$ and
$\langle \sigma \rangle =\sum_{n} \sigma_{n}/N$, over
the whole system.
The transition at the same temperature throughout the system is possibly
a characteristic of systems under the complete wetting (by ordered
phase) condition at the boundaries. 
In nematic liquid crystals, the phase transition of the whole system at
the same temperature has observed in systems under complete wetting
condition~\cite{Sheng1976};
while under the incomplete wetting condition these systems undergo the
``boundary-layer phase transition'' that occurs at a temperature higher
than the bulk transition temperature~\cite{Sheng1982}.

\begin{figure}
 \begin{center}
  \includegraphics[width=8.5cm,clip]{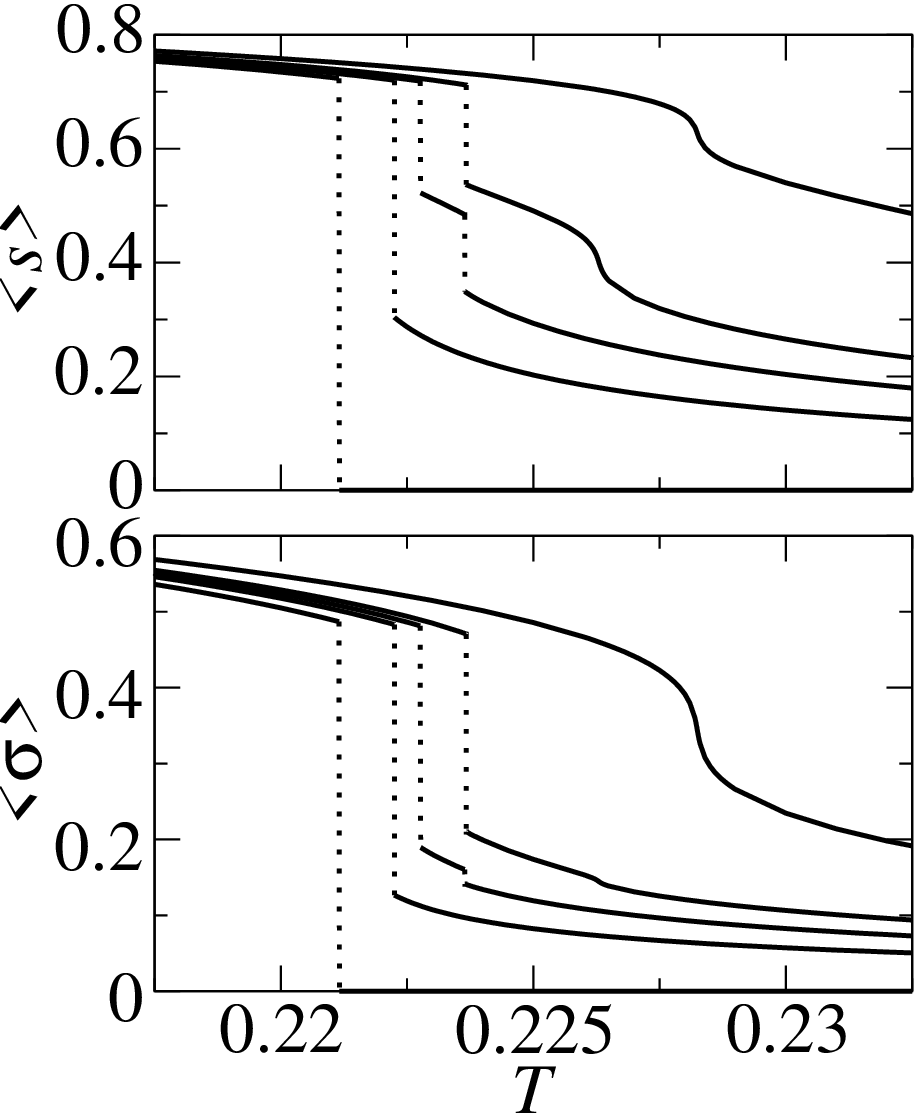}
 \end{center}
 \caption{
 Temperature dependence of order parameters $\langle s \rangle$ and
 $\langle \sigma \rangle$ for systems with $\alpha=1.0$.
 Curves correspond to the systems with thicknesses, reading from top to
 bottom, $N = 7, 13, 18, 25$, and infinity(i.e., bulk system). 
 }
 \label{fig:orderParameters}
\end{figure}

Figures~\ref{fig:orderParameters} show the temperature dependence of
order parameters $\langle s \rangle$ and $\langle \sigma \rangle$ for
$\alpha=1.0$.
These figures clearly exhibit three important features:
the increase in the transition temperatures, the existence of the
non-spontaneous N phase, and the existence of critical thicknesses below
which the phase transitions disappear.
The system with $\alpha=1.0$ exhibits direct I-A transition in the
absence of the walls.
By the influence of the walls the discontinuity at the transition
decreases but there is no qualitative difference in systems thicker
than $N=24$.
If the system becomes thinner than $N=24$, the N phase appears between I
and A phases.
Thus in systems with thickness $14 \leq N \leq 24$ three phases appear;
i.e., the non-spontaneous N phase appears in these systems.
As the system becomes thin, the discontinuities at the phase transitions
decreases and I-N transition vanishes at $N=13$;
eventually the I-A transition also vanishes at $N=7$.

\begin{figure}
 \begin{center}
  \includegraphics[width=8.5cm,clip]{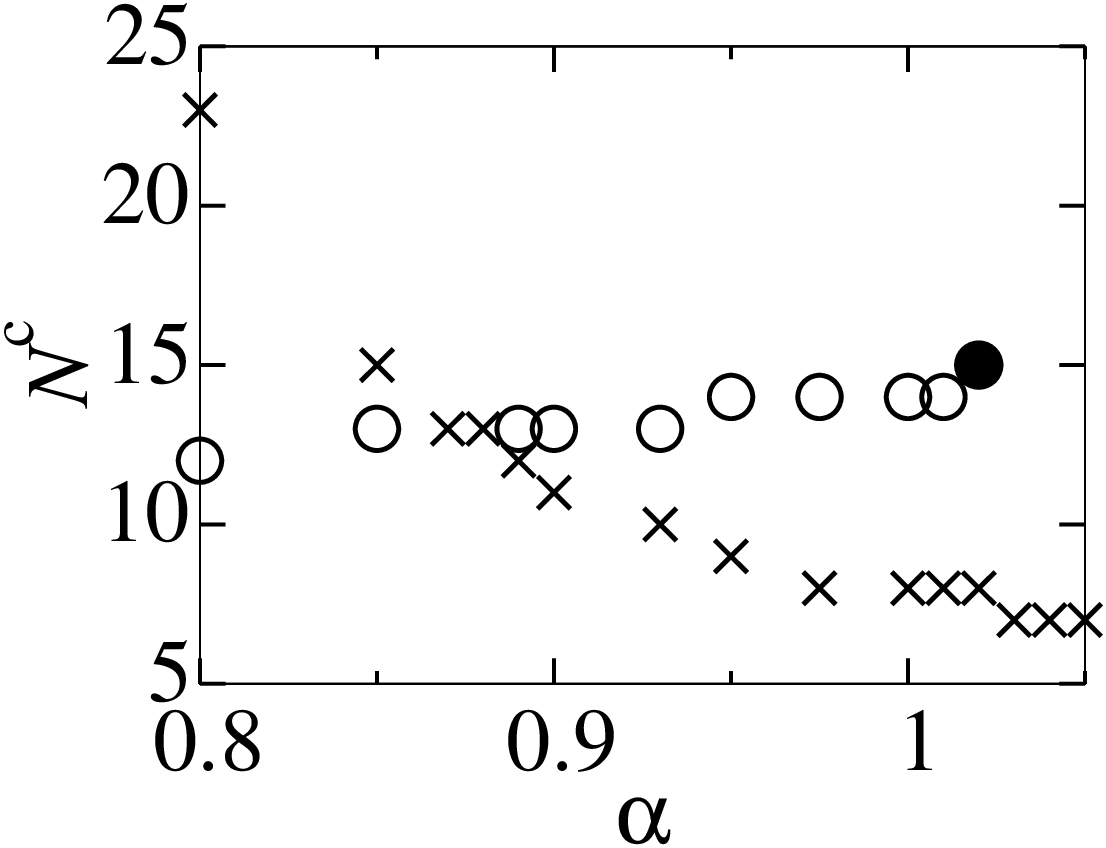}
 \end{center}
 \caption{
 The smectic A interaction parameter $\alpha$ dependence of critical
 thicknesses $N_\text{IN}^\text{c}$ and $N_\text{NA}^\text{c}$.
 The circles and  crosses denote $N_\text{IN}^\text{c}$ and
 $N_\text{NA}^\text{c}$, respectively.
 $N_\text{IN}^\text{c}(\alpha)$ terminates at $\alpha=1.02$ (indicated
 by a solid circle), since the nematic phase disappears if
 $\alpha > 1.02$. 
 }
 \label{fig:criticalN}
\end{figure}

%2
The critical thicknesses depend on the parameter $\alpha$;
the critical thicknesses for I-N and N-A phase transitions
($N_\text{IN}^\text{c}$ and $N_\text{NA}^\text{c}$, respectively) as
functions of $\alpha$ are shown in fig.~\ref{fig:criticalN}.
The $\alpha$-dependence of $N_\text{NA}^\text{c}$ is strong while that
of $N_\text{IN}^\text{c}$ is weak, since the parameter $\alpha$ directly
couples to the A phase order parameter in the one-particle potential 
\eqref{eq:oneParticlePotential}.
The N-A critical thickness seems to diverge as $\alpha$ approaches to
$0.70$ from above.
This is because the discontinuity at the N-A transition is small near
the tricritical point, and thus the boundary anchoring condition
strongly influences to the N-A transition.

%3
Figure~\ref{fig:criticalN} also shows that the curve
$N_\text{IN}^\text{c}$ terminates at $\alpha=1.02$, above which the
intermediate N phase never appears no matter what the thickness is and
only the I-A phase transition remains.
Since the N phase does not appear in the bulk systems with
$\alpha \geq \alpha_\text{triple} = 0.98$, the non-spontaneous N phase
appears in the systems with $\alpha_\text{triple} \leq \alpha \leq 1.02$.

Such a non-spontaneous N phase just above the $\alpha_\text{triple}$
corresponds to experimental results as follows.
The $\alpha$ vs temperature phase diagram for the McMillan model in the
absence of boundaries and fields can be compared with the alkyl chain
length vs temperature phase diagram because of the relation between
$\alpha$ and the molecular length~\cite{McMillan}.
These phase diagrams are very similar as shown in figs.2 and 6 of
ref.~\citen{McMillan}. 
The experimental result for 10CB liquid crystals sandwiched between
walls~\cite{Moses2001} shows that the non-spontaneous N phase appears at
the surface. 
Since the 10CB is the shortest molecule among members of the
homologous series of $n$CB exhibiting I-A transition, the 10CB can be
assumed to correspond to the material with the parameter $\alpha$ just
above the triple point $\alpha_\text{triple}$. 

\begin{figure}[ht]
 \begin{center}
  \includegraphics[width=8.5cm,clip]{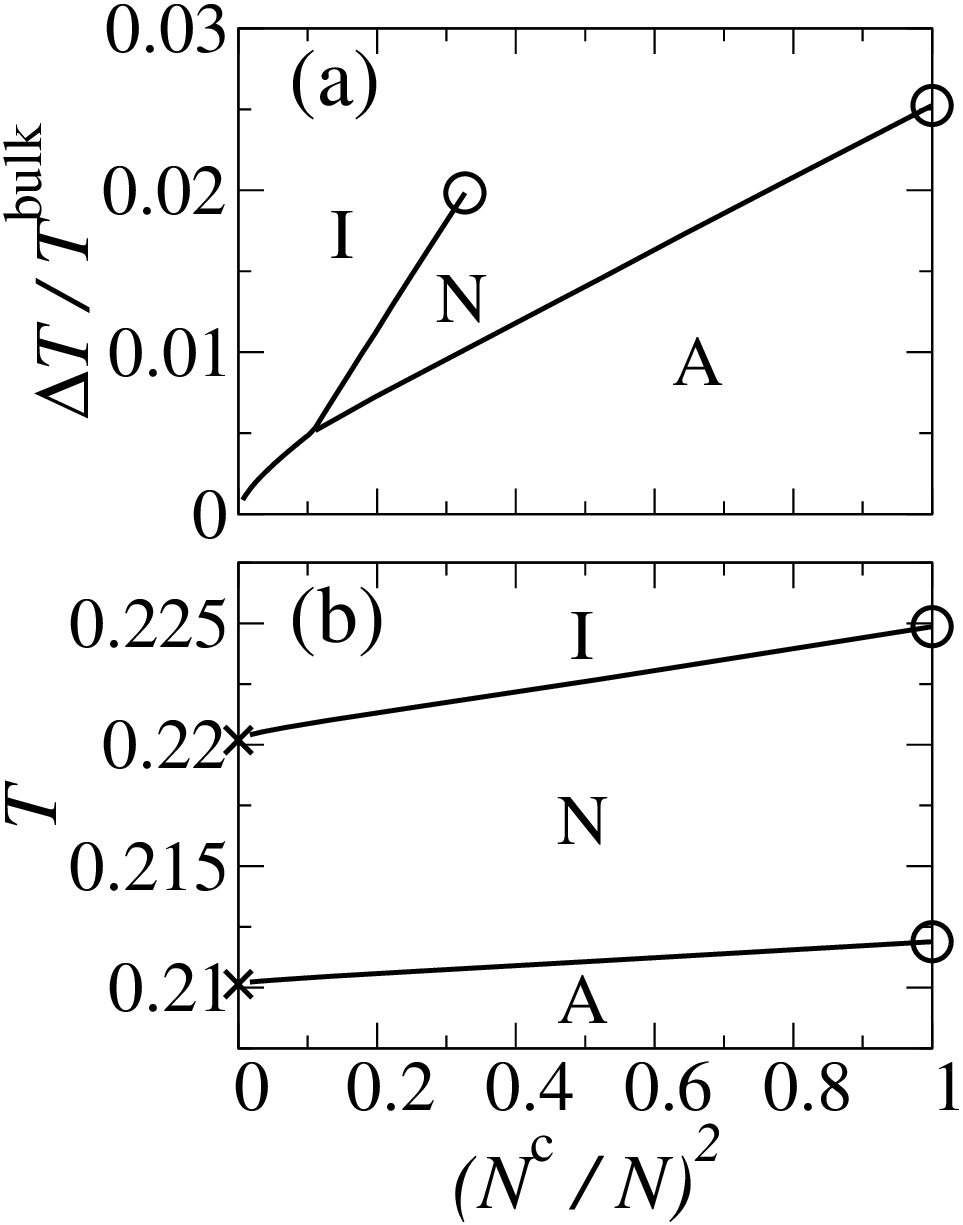}
 \end{center}
 \caption{
 Phase diagrams for (a)$\alpha=1.0$ and (b)$\alpha=0.88$ systems.
 (a) $T^\text{bulk}$ denotes the transition temperature without walls;
 $\Delta T$ is the change in transition temperature induced by anchoring
 walls. 
 The horizontal axis is normalized by the critical thickness
 $N^\text{c}=8$.
 The letters I, N, and A denote isotropic, nematic, and smectic A
 phases, respectively.
 Circles are the critical points.
 (b) The higher and lower curves denote I-N and N-A coexisting lines.
 The horizontal axis is normalized by the critical thickness
 $N^\text{c}=13$.
 Crosses denote the transition temperatures of systems without walls.
 }
 \label{fig:Ninv2vsT}
\end{figure}

\begin{figure}[ht]
 \begin{center}
  \includegraphics[width=8.5cm,clip]{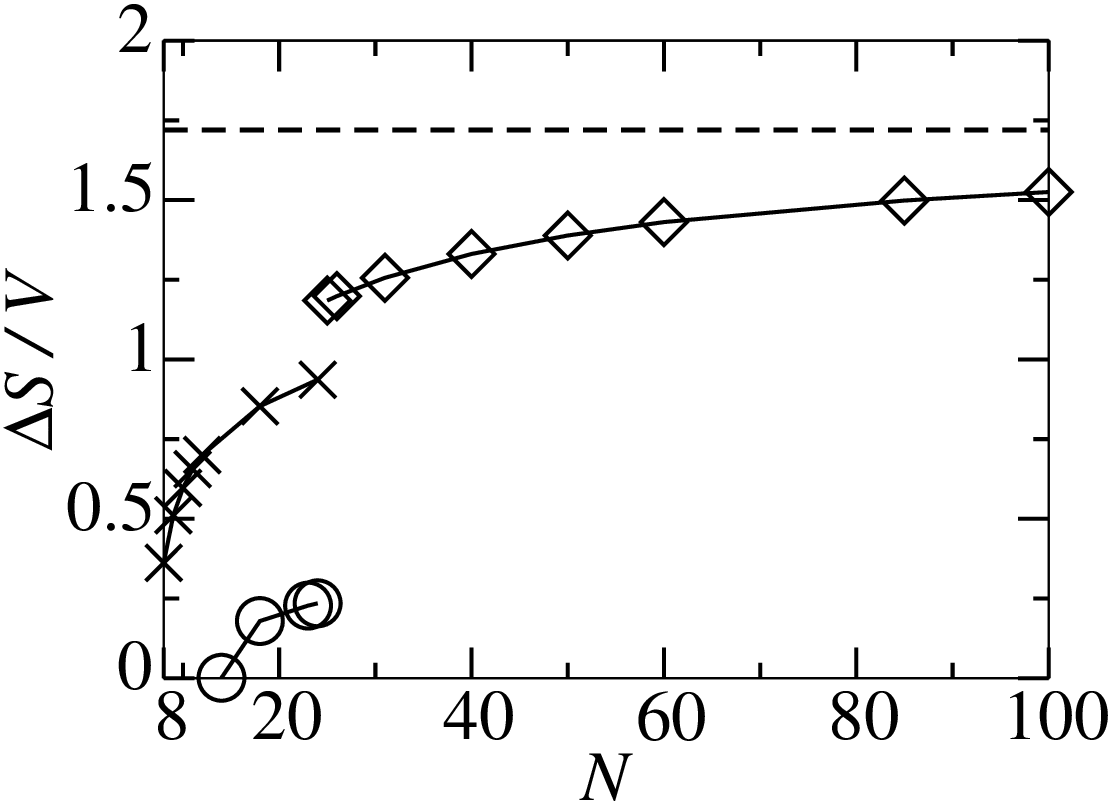}
 \end{center}
 \caption{
 Changes in entropy per volume at phase transitions of $\alpha=1.0$
 system. 
 The horizontal dashed line at $\Delta S/V = 1.72$ denotes the entropy
 change in the bulk system.
 The curves with $\circ$, $\times$, and $\diamond$ correspond to $\Delta
 S/V$ of I-N, N-A, and I-A phase transitions, respectively.
 }
 \label{fig:entropy}
\end{figure}

%4
The thickness dependences of the transition temperatures are shown in
figs.~\ref{fig:Ninv2vsT}(a) and (b) for $\alpha=1.0$ and $\alpha=0.88$,
respectively. 
From these figures, it seems that the increments of the transition
temperatures, $\Delta T(N) = T(N) - T^\text{bulk}$, depend on the system
thickness as $\Delta T \sim N^{-2}$ in a very thin region.
Such dependences differ from the prediction of the Kelvin
equation~\cite{Sluckin1990}:
\begin{equation}
 \Delta T(N) =
  \frac{2}{N} \frac{\gamma_\text{h}-\gamma_\text{l}}{L} T^\text{bulk},
  \label{eq:KelbinEq}
\end{equation}
where $L$ is the latent heat per volume;
$\gamma_\text{h}$ and $\gamma_\text{l}$ are, respectively, the surface
tensions (per a wall) of higher and lower temperature phases.
According to the Kelvin equation, the thickness dependence of
$\Delta T(N)$ is $\Delta T(N) \sim N^{-1}$.
This difference from the Kelvin equation clearly shows that the
assumption on which the Kelvin equation based is not valid in very thin
systems.
The Kelvin equation is derived by assuming that the latent heat and
surface tensions are independent of the system thickness, and this
assumption is accurate for sufficiently thick systems.
In fact as shown in fig.~\ref{fig:entropy}, the entropy of transition
(i.e., $L/T$) strongly depends on the thickness in thin region;
the Kelvin equation is valid for much thicker systems than the system we
considered here.
The Clapeyron-Clausius equation,
which is valid irrespective of the system
thickness~\cite{Poniewierski1987}, is appropriate to treat such thin
systems.

\section{Summary and Discussions}\label{Discussion}
%1
In summary, we have investigated the influence of the strongly
homeotropic anchoring walls on the McMillan model system.
We have shown that the existence of the critical thickness for I-N, N-A
and I-A transitions.
This result is analogous to the existence of critical thickness for I-N
transition in Landau-de Gennes nematogen model~\cite{Sheng1976}.
We have found that the system with $\alpha$ just above the
$\alpha_\text{triple}$ exhibits non-spontaneous N phase induced by
anchoring walls;
such a behavior is consistent with the behavior of a homologous series
$n$CB~\cite{Moses2001}. 
%We have found the non-spontaneous N phase induced by anchoring
%walls~\cite{Lelidis2002}.
We have also obtained the I-N, N-A and I-A transition temperatures as
functions of the system thickness.
We have confirmed that the Kelvin equation is not valid for very thin
system;
the transition temperature behaves as $T(N)-T^\text{bulk} \sim N^{-2}$
instead of the prediction of the Kelvin equation
$T(N)-T^\text{bulk} \sim N^{-1}$.

%2
We emphasize that the effects of anchoring walls are qualitatively
similar to the effects of the external field as indicated in the early studies~\cite{Sheng1976,Poniewierski1987}.
Let us introduce two external fields $h_{s}$ and $h_{\sigma}$,
conjugate to the order parameters $s$ and $\sigma$, respectively.
We note that the squares of electric and magnetic fields correspond to
$h_{s}$, while $h_{\sigma}$ has no experimental counterpart;
thus usually only the field $h_s$ has been considered in
literatures~\cite{Wojtowicz1974,Rosenblatt1981,Hama,Lelidis1993a,Lelidis1993b,Lelidis1994,Lelidis1996,Galatora1998},
and $h_{\sigma}$ has been omitted.
In the following, we first consider only the influence of the field
$h_{s}$.
\begin{figure}[b]
 \begin{center}
  \includegraphics[width=8.5cm,clip]{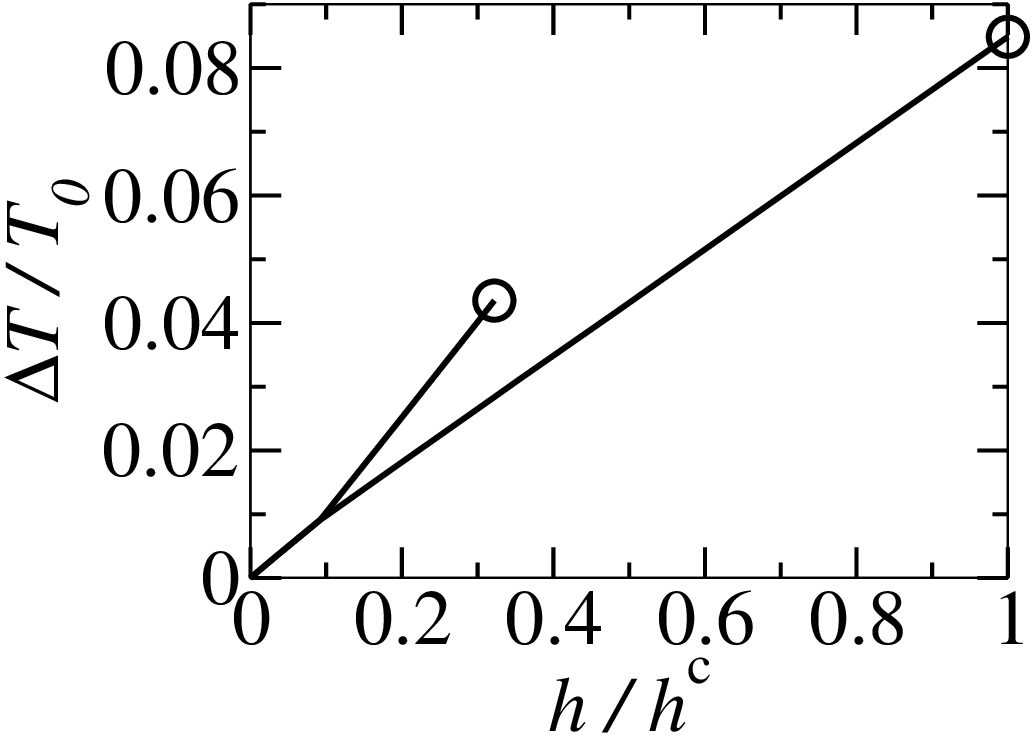}
 \end{center}
 \caption{
 The transition temperatures $T(h_{s}, h_{\sigma})$ for $\alpha = 1.0$
 system as a function of external fields $h_{s}$ and $h_{\sigma}$.
 This figure shows field-dependence of $\Delta T/T_{0}$, where
 $\Delta T = T(h_{s}, h_{\sigma})-T(0, 0)$ and $T_{0} = T(0,0)$ is the 
 transition temperature in the absence of external fields;
 this figure is restricted to a plane $h = h_{s} = h_{\sigma}$ for
 simplicity. 
 The curve in $0 \leq h/h^\text{c} \lesssim 0.1$ denotes I-A transition
 temperature; higher branch and lower branch in
 $h/h^\text{c} \gtrsim 0.1$ denote I-N and N-A transition temperatures,
 respectively.
 Circles indicate the critical points of the corresponding phase
 transitions. 
 }
 \label{fig:hsVsT}
\end{figure}
The increase of the transition temperature and the non-spontaneous N
phase occur under the influence of $h_{s}$;
in this sense, the influence of the field $h_{s}$ and of the anchoring
walls are qualitatively similar.
However, the similarity is incomplete with respect to 
the existence of the critical points.
By the application of sufficiently strong field $h_{s}$ the I-N
transition vanishes but the N-A transition does not, i.e., the critical
point on the N-A coexisting line of $h_{s}$-$T$ phase diagram does not
exist. 
The N-A transition in a bulk McMillan system changes from first-order to
second-order transition by the influence of $h_{s}$;
i.e. the tricritical point appears on the N-A coexisting line instead of
the critical point~\cite{Lelidis1994,Lelidis1996,Hama}.
Thus we can conclude that the similarity is broken since both the I-N
and N-A coexisting line terminate at critical points in the system under
the influence of the anchoring walls, as shown in
fig.~\ref{fig:Ninv2vsT}. 
However, if we consider the external field $h_{\sigma}$ in addition to
$h_{s}$, the tricritical point on the N-A coexisting line changes to a
critical point in the bulk system~\cite{Torikai2004};
then the similarity is again recovered.
In order to show this similarity explicitly, we calculate the increment
in transition temperatures under the application of both the $h_{s}$ and
$h_{\sigma}$ (see ref.~\cite{Torikai2004} for details of the transition
temperatures under the external fields).
We plot on fig.~\ref{fig:hsVsT} the increment of the transition
temperature $\Delta T = T(h_{s}, h_{\sigma})-T(0, 0)$, under a condition
$h_{s}=h_{\sigma}$ for simplicity.
The similarity between the influence of anchoring walls and of external
fields is clearly seen in fig.~\ref{fig:Ninv2vsT} and
fig.~\ref{fig:hsVsT}. 
This fact shows that, in considering the anchoring effects in the
relationship with the external field, it is necessary to introduce the
fictitious field $h_{\sigma}$ together with $h_{s}$.

\end{document}